# Conversion of linearly polarized high-harmonic radiation into the circularly polarized in an optically modulated plasma-based X-ray laser: taking into account the nonlinearity and finite width of the harmonic spectral lines

V. A. Antonov[1], I. R. Khairulin[1,2], and M. Yu. Ryabikin[1,2]

[1] *Gaponov-Grekhov Institute of Applied Physics of the Russian Academy of Sciences,*
*46 Ulyanov Street, Nizhny Novgorod 603950, Russia*
[2]*Lobachevsky State University of Nizhny Novgorod,*
*23 Gagarin Avenue, Nizhny Novgorod 603950, Russia*

Corresponding author: V. A. Antonov, antonov@appl.sci-nnov.ru

The method proposed in [I.R. Khairulin et al., JETP Letters 117(9), 652 (2023)] for converting the linearly polarized radiation of high-order harmonics of an optical field into the circularly polarized in an optically modulated active medium of a plasma-based X-ray laser has been developed. This method is extended to the case of finite width of spectral lines (finite pulse duration) of high harmonic radiation and nonlinear interaction of radiation with matter. The transformation of the polarization of the harmonic field is achieved by introducing a phase shift of $\pi/2$ between its polarization components, parallel and orthogonal to the polarization of the modulating field, through resonant dispersion on the spectral wings of the induced gain lines of the active medium and is accompanied by an increase in the power of high harmonics. It is shown that for harmonics with a finite bandwidth, the polarization of the output radiation is non-uniform in time. The reasons for such non-uniformity are analyzed and the conditions for its minimization are found. The influence of nonlinearity and amplified spontaneous radiation of the active medium on the high-harmonic polarization is investigated. The possibilities for the experimental implementation of this effect in a neon-like active medium of a plasma X-ray laser based on $Ti^{12+}$ ions with an unperturbed wavelength of the inverted transition at 32.6 nm are analyzed.

## 1. Introduction

Circularly polarized radiation of extreme ultraviolet (XUV) and X-ray ranges is a powerful tool for studying magnetic and chiral media exhibiting circular dichroism at the boundaries of absorption bands from inner electron shells [1-8]. Until recently, synchrotron sources were predominantly used to generate circularly polarized XUV/X-ray radiation [1-4]. However, the picosecond duration of synchrotron radiation limits the possibilities of its application for studying the dynamics of ultrafast physical and chemical processes. This limitation was partly overcome with the advent of free-electron X-ray lasers, which make it possible to generate femtosecond XUV/X-ray radiation with circular polarization and sufficiently high power [9-13]. However, the stochastic nature of the generated radiation (even in the self-seeding regime) and the limited availability of the X-ray free-electron lasers complicate their practical use.

The shortest duration (down to tens of attoseconds) and the highest coherence of XUV/X-ray radiation are provided by sources based on the generation of high-order harmonics of the optical field (HHG) [14-16, 5-8]. As a rule, atomic gases irradiated with a femtosecond laser field of the visible or infrared (IR) range with an intensity of about $10^{14}$–$10^{15}$ W/cm$^2$ are used for HHG. However, in a single-frequency laser field, this approach does not allow efficient generation of XUV/X-ray radiation with circular or close to circular polarization [17, 18] due to the rapid decrease in the harmonic yield with increasing ellipticity of the laser field. One of the rare exceptions is harmonics that are resonant with one of the quantum transitions of the medium in which the

HHG occurs [5]. However, due to the narrowness of the resonances and the non-equidistant positions of their frequencies, this is observed only in a narrow spectral range, not exceeding several eV. At the same time, when a two-frequency laser field is used (usually at the fundamental frequency and its second harmonic) with different polarizations of the spectral components, the polarization of the generated high-order harmonics becomes spectrally non-uniform (for example, adjacent harmonics can have alternating right and left circular polarization) [6-8, 19-20]. This obstacle can be circumvented by using a gas of aligned molecules instead of atoms. However, with HHG in a circularly polarized laser field, this approach is characterized by low efficiency, two orders of magnitude lower than the efficiency of HHG in a linearly polarized field [21]. On the other hand, if a linearly polarized laser field is used, the ellipticity of the high-order harmonics generated in a gas of aligned molecules turns out to not exceed 0.4 [22]. Another alternative approach is to convert the high harmonic radiation from linearly polarized to circularly polarized using phase-shifting X-ray optics such as mirrors or multilayer quarter-wave plates [23, 24]. However, this approach significantly (by two orders of magnitude) reduces the energy of the converted radiation and operates in a relatively narrow spectral range with a width of the order of several eV.

Thus, the following tasks are on the agenda: (a) increasing the power of elliptically and circularly polarized high-order harmonics and (b) efficient conversion of linearly polarized high-harmonic radiation into circularly polarized one. In [25], the amplification of circularly polarized radiation of a single high-order harmonic with a wavelength of 32.8 nm in the active medium of a plasma X-ray laser based on nickel-like $Kr^{8+}$ ions was experimentally demonstrated. At the same time, the small bandwidth of the gain line of the plasma-based X-ray laser, which is about $10^{-5}$ of the frequency of the amplified radiation and about $10^{-3}$ of the frequency of the optical field used for the HHG, prevents the simultaneous amplification of harmonics of different orders. Moreover, in the process of amplification, the spectrum of the field of the amplified harmonic narrows. As a result, the duration of the amplified radiation increases to fractions of picoseconds, which prevents its use for studying and controlling the femto- and attosecond processes.

To circumvent this limitation, it was proposed to irradiate the active medium of a plasma-based X-ray laser with a replica of the IR field of the fundamental frequency used for HHG [26], the role of which in this case is to modulate the frequency of the inverted transition of the plasma-based X-ray laser at a doubled optical frequency by means of the quadratic Stark effect (its time-variable component). This optical modulation of the inverted transition frequency leads to a redistribution of the gain of the active medium of the X-ray laser over combination frequencies separated by a double frequency of the optical field, which makes it possible to amplify a set of harmonics of different orders of the modulating field. This approach potentially allows increasing the energy of subfemtosecond pulses of circularly polarized high-order harmonics by tens of times. At the same time, the anisotropy of the gain for the polarization components of the harmonic field, parallel and orthogonal to the polarization of the modulating field, allows increasing the ellipticity of the harmonic radiation in the amplification process by several times.

In a recent paper [27], a version of this approach was proposed that allows the linearly polarized high-harmonic radiation to be converted into circularly polarized radiation with a simultaneous increase in power. To achieve this goal, it was proposed to use the anisotropy of the resonant dispersion of the active medium induced by the modulating field and to impart a phase difference of π/2 between the polarization components of the harmonic field, parallel and orthogonal to the polarization of the modulating field, as a result of their interaction with the tails of the induced gain lines. To maximize the anisotropy of the resonance dispersion, it was proposed (by appropriately choosing the intensity of the IR field) to detune the induced gain lines for the orthogonal polarization components of the harmonic field from each other and to use the frequency interval between them for the transformation. The fundamental possibility of such a polarization transformation was shown in [27] in the approximations of (a) monochromatic fields (neglecting the finite bandwidth of the harmonics) and (b) linear interaction of the harmonic radiation with the active medium. In addition, the difference in gain factors for orthogonal polarization components of the field of each harmonic was not taken into account. This limits the possibilities of using the results of the above-mentioned work for planning an experiment on transforming the polarization state of the field of high harmonics in the process of its amplification.

The aim of this work is to generalize the results of the article [27] to a case closer to the experiment of a high-harmonic field with a finite bandwidth (i.e. with a finite duration of the envelope), taking into account also the nonlinearity in the interaction of radiation with the active medium, its amplified spontaneous emission, as well as the difference in the gain factors between the polarization components of the harmonic field. It is shown that the possibility of polarization transformation of high harmonic radiation is preserved even when these factors are taken into account; however, the ellipticity of the radiation becomes non-uniform on the scale of its envelope, and the optimal conditions for polarization transformation change. The influence of the parameters of the seed radiation, the modulating field, and the active medium on the course of changes in the ellipticity of the output radiation over time is investigated, and the conditions that reduce these changes to a minimum are analyzed. The paper presents estimates of the experimental parameters required for the conversion of high-harmonic polarization in the active medium of a plasma X-ray laser based on neon-like $Ti^{12+}$ ions with unperturbed gain at a wavelength of 32.6 nm, irradiated by an IR laser field with a wavelength of 3.9 μm.

The paper is organized as follows. Introduction (Sec. I) is followed by a description of the theoretical model and the equations used (Sec. II); then an analytical solution and a discussion of the conditions for transforming the polarization of high-harmonic radiation (Sec. III) are presented. The subsequent sections analyze the optimal conditions for transforming the polarization of a single high-order harmonic radiation (Sec. IV), as well as a set of such harmonics (Sec. V). Section VI summarizes the main findings and concludes the paper.

**2. Theoretical model**

This work is a generalization of the paper [27], and is based on a similar theoretical model. In this regard, up to equation (5), this section largely reproduces the similar section in the above-mentioned article.

Let us consider the active medium of a plasma X-ray laser with a population inversion at the transition $3p^1S_0\rangle\leftrightarrow|3p^1P_1\rangle$ between two excited energy levels of neon-like $Ti^{12+}$ ions with an unperturbed transition wavelength of 32.6 nm [28, 29]. The upper energy level is non-degenerate and corresponds to the state $|1\rangle = |3p^1S_0, J = 0, M = 0\rangle$ with the total angular momentum $J = 0$. The lower energy level is triply degenerate and corresponds to the states $|2\rangle = |3p^1P_1, J = 1, M = 0\rangle$, $|3\rangle = |3p^1P_1, J = 1, M = 1\rangle$, and $|4\rangle = |3p^1P_1, J = 1, M = -1\rangle$ with the total angular momentum $J = 1$ and the projections of the angular momentum onto the quantization axis $M = 0, \pm 1$. Under typical experimental conditions, the active medium has the shape of a thin cylinder elongated along the $x$ axis with a concentration of $Ti^{12+}$ ions and free electrons $N_{ion} = 4.2\times10^{18}$ cm$^{-3}$ and $N_e = 5\times10^{19}$ cm$^{-3}$, respectively (see [26, 27]). In the interaction region, only about 1% of the ions are in the upper state $|1\rangle$, while the lower states $|2\rangle$, $|3\rangle$, and $|4\rangle$ are not populated due to fast radiative transitions to even lower energy levels of $Ti^{12+}$. As a result, the intensity gain of the active medium in the region under consideration is 70 cm$^{-1}$.

Further we will assume that the active medium of the X-ray laser is simultaneously irradiated by the IR field of frequency $\Omega$ and the radiation of its high harmonics. Both fields are linearly polarized and propagate along the $x$ axis of the plasma channel, see Fig. 1. We will assume that the duration of the IR field pulse significantly exceeds the characteristic time scales of the processes under consideration, and the amplitude of the IR field in the region of interaction with the radiation of harmonics changes insignificantly.

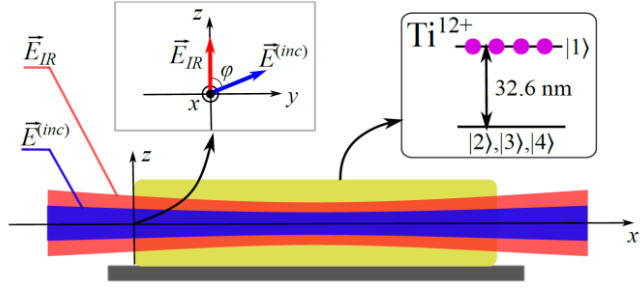

**Fig. 1.** Schematic representation of the proposed experiment. The pre-created active medium of the plasma X-ray laser based on neon-like Ti$^{12+}$ ions (yellow region) is simultaneously irradiated by a linearly polarized IR laser field (red region) and a set of linearly polarized high harmonics of the IR field (blue region), forming a train of subfemtosecond pulses. The left inset shows the mutual orientation of the IR-field and high-harmonic field polarizations (red and blue arrows, respectively). The right inset shows the diagram of unperturbed resonant energy levels of Ti$^{12+}$ ions.

In the case described above, IR radiation can be represented as a plane monochromatic wave, the electric field of which is given by

$$\vec{E}_{IR}(x,t) = \vec{z}_0 E_{IR} \cos\left[\Omega\left(t - x\sqrt{\varepsilon_{pl}^{(IR)}}/c\right)\right], \quad (1)$$

where $\vec{z}_0$ is the unit vector of polarization of the laser field (hereinafter the $z$ axis will be chosen as the quantization axis), $E_{IR}$ is the amplitude of the electric field, $c$ is the speed of light in a vacuum, $\varepsilon_{pl}^{(IR)} = 1 - \omega_{pl}^2/\Omega^2$ is the permittivity of the plasma at the frequency of the IR field, $\omega_{pl} = \sqrt{4\pi N_e e^2/m_e}$ is the plasma frequency, $e$ and $m_e$ are the charge and mass of the electron, respectively. In Eq. (1), the values of $\Omega$ and $E_{IR}$ are such that the IR field frequency and the Rabi frequencies at all dipole-allowed transitions from the $|1\rangle$ -$|4\rangle$ states are much lower than the frequencies of these transitions. Thus, the influence of the field (1) is mainly reduced to a change in the instantaneous values of the energy of the states $|1\rangle$ - $|4\rangle$ in time and space on a scale of a fraction of the optical cycle and wavelength due to the quadratic Stark effect [30]. The frequencies of transitions between the states $|1\rangle$ - $|4\rangle$ change following a change in the square of the modulus of the laser field (1):

$$\omega_{12}(t,x) = \bar{\omega}_z + \Delta_z \cos\left[2\Omega\left(t - x\sqrt{\varepsilon_{pl}^{(IR)}}/c\right)\right], \; \omega_{13}(t,x) = \omega_{14}(t,x) = \bar{\omega}_y + \Delta_y \cos\left[2\Omega\left(t - x\sqrt{\varepsilon_{pl}^{(IR)}}/c\right)\right],$$
$$\omega_{23}(t,x) = \omega_{24}(t,x) = (\Delta_z - \Delta_y)\left\{1 + \cos\left[2\Omega\left(t - x\sqrt{\varepsilon_{pl}^{(IR)}}/c\right)\right]\right\}, \; \omega_{34}(t,x) = 0, \quad (2)$$

where $\Delta_z \equiv \Delta_{12}$, $\Delta_y \equiv \Delta_{13} = \Delta_{14}$, $\Delta_{ij} = \sum_{k \neq i}(|d_{ki}^{(z)}|E_{IR})^2/(2\hbar^2\omega_{ik}) - \sum_{k \neq j}(|d_{kj}^{(z)}|E_{IR})^2/(2\hbar^2\omega_{jk})$ is the amplitude of the modulation of the $|i\rangle \to |j\rangle$ transition frequency, $d_{ki}^{(z)}$ is the $z$ projection of the dipole moment of the transition from the state $|i\rangle$ to the state $|k\rangle$ with frequency $\omega_{ik}$, $\hbar$ is the reduced Planck constant, and the summation is carried out over all states of the Ti$^{12+}$ ion in the absence of a field (1). Also in (2) $\bar{\omega}_z \equiv \bar{\omega}_{12} = \omega_{12}^{(0)} + \Delta_z$ is the time-average frequency of the transition $|1\rangle \to |2\rangle$ and $\bar{\omega}_y \equiv \bar{\omega}_{13} = \omega_{13}^{(0)} + \Delta_y = \bar{\omega}_{14} = \omega_{14}^{(0)} + \Delta_y$ is the time-average frequency of the transitions $|1\rangle \to |3\rangle$ and $|1\rangle \to |4\rangle$. Note that for neon-like Ti$^{12+}$ ions, the modulation amplitudes of the transitions $|1\rangle \to |2\rangle$ and $|1\rangle \to |3\rangle, |4\rangle$ differ slightly: $|\Delta_z|/|\Delta_y| \approx 0.93$, which leads to a difference in the time-averaged frequencies of the corresponding transitions.

Simultaneously with the IR field of fundamental frequency $\Omega$, the active medium is irradiated by a set of its $2N+1$ high harmonics, linearly polarized in the $zy$ plane. The amplitudes and initial phases of the harmonics are assumed to be equal. Accordingly, in the time domain, the radiation of harmonics forms a train of subfemtosecond pulses, the electric field of which has the form

$$\vec{E}^{(inc)}(t) = \frac{1}{2}\left[\vec{z}_0 \cos(\varphi) + \vec{y}_0 \sin(\varphi)\right] E_0 a(t) \exp(-i\omega_{inc}t) \sum_{n=-N}^{N} e^{-i2n\Omega t} + \text{c.c.}, \qquad (3)$$

where $E_0$ is the amplitude of each harmonic, $\omega_{inc}$ is the carrier frequency of field (3), $\varphi$ is the angle between the directions of polarization of the high-harmonic field and the laser field (1) (see Fig. 1), and $a(t)$ is the field envelope of the pulse train, which determines the shape of the spectral contour of each individual harmonic. We will assume that the carrier frequency $\omega_{inc}$ of field (3) is close to the average frequencies $\bar{\omega}_z$ and $\bar{\omega}_y$. Then, when propagating in the medium, field (3) will lead to the excitation of resonant polarization, which is determined as follows:

$$\vec{P}(x,t) = N_{ion}^{(res)}\left(\rho_{12}\vec{d}_{21} + \rho_{13}\vec{d}_{31} + \rho_{14}\vec{d}_{41}\right) + \text{c.c.}, \qquad (4)$$

where $N_{ion}^{(res)}$ is the concentration of $Ti^{12+}$ ions, which are in one of the resonance states $|1\rangle$, $|2\rangle$, $|3\rangle$ or $|4\rangle$ before exposure to high-harmonic radiation (3); $\rho_{12}$ is the quantum coherence (off-diagonal element of the density matrix) at the transition $|1\rangle \to |2\rangle$, $\rho_{13}$ and $\rho_{14}$ are the quantum coherences at the transitions $|1\rangle \to |3\rangle, |4\rangle$, respectively; $\vec{d}_{21} = \vec{z}_0 d_z$, $\vec{d}_{31} = \vec{d}_{41} = -i\vec{y}_0 d_y$, $d_z = D/\sqrt{3}$, $d_y = D/\sqrt{6}$, and $D \equiv \left|\langle 3p^1S_0 \| D \| 3s^1P_1\rangle\right| \simeq 0.41$ at.un. is the reduced dipole moment of the inverted transition. Note that the dipole moment of the transition $|1\rangle \to |2\rangle$ is oriented along the $z$ quantization axis, and the dipole moments of the transitions $|1\rangle \to |3\rangle, |4\rangle$ lie in the $xy$ plane. Thus, the transition $|1\rangle \to |2\rangle$ leads to an amplification of the $z$-polarized component of the harmonic radiation (3), and the transitions $|1\rangle \to |3\rangle, |4\rangle$ lead to an amplification of its $y$-polarized component. Further, for brevity, we will call the transitions $|1\rangle \to |2\rangle$ and $|1\rangle \to |3\rangle, |4\rangle$ $z$-polarized and $y$-polarized transitions, respectively.

The propagation of radiation (3) in an active plasma medium modulated by an optical field will be described below by a spatially one-dimensional wave equation, neglecting the change in the characteristics of the harmonic field in the transverse direction:

$$\frac{\partial^2 \vec{E}}{\partial x^2} - \frac{\varepsilon_{pl}^{(XUV)}}{c^2}\frac{\partial^2 \vec{E}}{\partial t^2} = \frac{4\pi}{c}\frac{\partial^2 \vec{P}}{\partial t^2}, \qquad (5)$$

where $\vec{E}$ is the electric field vector of the harmonics in the medium, and $\varepsilon_{pl}^{(XUV)} = 1 - \omega_{pl}^2/\omega_{inc}^2$ is the permittivity of the plasma at the carrier frequency of radiation (3). The evolution of the quantum state of resonant $Ti^{12+}$ ions is described here by equations for the elements of the density matrix $\rho_{ij}$:

$$\begin{cases} \dfrac{\partial \rho_{11}}{\partial t} + \gamma_{11}\rho_{11} = \dfrac{i}{\hbar}\sum_{s=1}^{4}\left(\rho_{s1}\vec{d}_{1s} - \rho_{1s}\vec{d}_{s1}\right)\vec{E}, \\[2mm] \dfrac{\partial \rho_{ii}}{\partial t} + \gamma_{ii}\rho_{ii} = A\rho_{11} + \dfrac{i}{\hbar}\sum_{s=1}^{4}\left(\rho_{si}\vec{d}_{is} - \rho_{is}\vec{d}_{si}\right)\vec{E}, \quad i \ne 1, \\[2mm] \dfrac{\partial \rho_{ij}}{\partial t} + \left(i\omega_{ij}(t,x) + \gamma_{ij}\right)\rho_{ij} = \dfrac{i}{\hbar}\sum_{s=1}^{4}\left(\rho_{sj}\vec{d}_{is} - \rho_{is}\vec{d}_{sj}\right)\vec{E}, \quad j \ne i, \end{cases} \qquad (6)$$

where the spatiotemporal dependences of the $|i\rangle \leftrightarrow |j\rangle$ transition frequencies are determined by equations (2), $A$ is the rate of spontaneous radiative transitions from the state $|1\rangle$ to each of the states $|2\rangle$, $|3\rangle$, and $|4\rangle$, $1/A = 242.5$ ps [31], and $\gamma_{ij}$ are the relaxation rates of the density matrix elements, determined as follows. The relaxation rate of the diagonal element of the density matrix $\rho_{ii}$ is the sum of the rates of radiative transitions from the state $|i\rangle$ to all lower states, $\Gamma_{rad}^{(i)}$, and the rate of tunnel ionization from the state $|i\rangle$ under the action of the IR field, $w_{ion}^{(i)}$, calculated using the Perelomov-Popov-Terentyev formula [32], i.e. $\gamma_{ii} = \Gamma_{rad}^{(i)} + w_{ion}^{(i)}$. Note that for the IR field intensities considered below, ionization from the resonant states is insignificant, $w_{ion}^{(i)} \ll \Gamma_{rad}^{(i)}$ ($\forall i$). At the same time, the relaxation rate of the off-diagonal element of the density matrix $\rho_{ij}$ has the form

$\gamma_{ij} = (\gamma_{ii}+\gamma_{ij})/2+\gamma_{Coll}$, where $\gamma_{Coll}$ is the collision frequency in the plasma. The value of $\gamma_{Coll}$ was calculated from the experimentally measured width of the spectral gain line of an optically thin medium (see [26]) and is $1/\gamma_{Coll} = 213$ fs.

Next, we will move on to local time $t \to \tau = t - x\sqrt{\varepsilon_{pl}^{(XUV)}}/c$ and will seek a solution to the system of equations (2), (4)-(6) in the slowly varying amplitude approximation, assuming

$$\vec{F}(x,\tau) = \frac{1}{2}\left(\vec{z}_0 \tilde{F}_z(x,\tau) + \vec{y}_0 \tilde{F}_y(x,\tau)\right)\exp(-i\omega_{inc}\tau) + \text{c.c.},$$

$$\rho_{12}(x,\tau) = \tilde{\rho}_{12}(x,\tau)e^{-i\omega_{inc}\tau},\ \rho_{13}(x,\tau) = \tilde{\rho}_{13}(x,\tau)e^{-i\omega_{inc}\tau},\ \rho_{14}(x,\tau) = \tilde{\rho}_{14}(x,\tau)e^{-i\omega_{inc}\tau},$$

$$\rho_{ij}(x,\tau) = \tilde{\rho}_{ij}(x,\tau),\ ij \ne \{12,21,13,31,14,41\},$$

$$\rho_{ij} = \rho_{ji}^*,$$

(7)

where $F$ denotes $E$ or $P$; $\tilde{F}_z(x,\tau)$ and $\tilde{F}_y(x,\tau)$ correspond to the slowly varying complex amplitudes of the polarization components of the resonant XUV radiation or the resonant polarization of the medium, $|\partial \tilde{F}_{z,y}/\partial \tau| \ll \omega_{inc}|\tilde{F}_{z,y}|$ and $|\partial \tilde{F}_{z,y}/\partial x| \ll \omega_{inc}\sqrt{\varepsilon_{pl}^{(XUV)}}|\tilde{F}_{z,y}|/c$, while $\tilde{\rho}_{ij}(x,\tau)$ denote the slowly varying amplitudes of the elements of the density matrix of the medium, $|\partial \tilde{\rho}_{ij}/\partial \tau| \ll \omega_{inc}|\tilde{\rho}_{ij}|$ and $|\partial \tilde{\rho}_{ij}/\partial x| \ll \omega_{inc}\sqrt{\varepsilon_{pl}^{(XUV)}}|\tilde{\rho}_{ij}|/c$ and $|\partial \tilde{F}_{z,y}/\partial x| \ll \omega_{inc}\sqrt{\varepsilon_{pl}^{(XUV)}}|\tilde{F}_{z,y}|/c$. These simplifications reduce the wave equation (5) to the form

$$\frac{\partial \vec{\tilde{E}}}{\partial x} = i2\pi \frac{\omega_{inc}}{c\sqrt{\varepsilon_{pl}^{(XUV)}}} \vec{\tilde{P}}, \tag{8}$$

while the equations for the elements of the density matrix take the form

$$\begin{cases}
\dfrac{\partial \tilde{\rho}_{11}}{\partial \tau} + \gamma_{11}\tilde{\rho}_{11} = \dfrac{i}{2\hbar}d_z\left(\tilde{\rho}_{12}^*\tilde{E}_z - \tilde{\rho}_{12}\tilde{E}_z^*\right) - \dfrac{1}{2\hbar}d_y\left(\tilde{\rho}_{13}^*\tilde{E}_y + \tilde{\rho}_{13}\tilde{E}_y^*\right) - \dfrac{1}{2\hbar}d_y\left(\tilde{\rho}_{14}^*\tilde{E}_y + \tilde{\rho}_{14}\tilde{E}_y^*\right), \\[4pt]
\dfrac{\partial \tilde{\rho}_{22}}{\partial \tau} + \gamma_{22}\tilde{\rho}_{22} = A\tilde{\rho}_{11} - \dfrac{i}{2\hbar}d_z\left(\tilde{\rho}_{12}^*\tilde{E}_z - \tilde{\rho}_{12}\tilde{E}_z^*\right), \\[4pt]
\dfrac{\partial \tilde{\rho}_{33}}{\partial \tau} + \gamma_{33}\tilde{\rho}_{33} = A\tilde{\rho}_{11} + \dfrac{1}{2\hbar}d_y\left(\tilde{\rho}_{13}^*\tilde{E}_y + \tilde{\rho}_{13}\tilde{E}_y^*\right), \\[4pt]
\dfrac{\partial \tilde{\rho}_{44}}{\partial \tau} + \gamma_{44}\tilde{\rho}_{44} = A\tilde{\rho}_{11} + \dfrac{1}{2\hbar}d_y\left(\tilde{\rho}_{14}^*\tilde{E}_y + \tilde{\rho}_{14}\tilde{E}_y^*\right), \\[4pt]
\dfrac{\partial \tilde{\rho}_{12}}{\partial \tau} + \left[i\left(\omega_{12}(\tau,x) - \omega_{inc}\right) + \gamma_{12}\right]\tilde{\rho}_{12} = -\dfrac{i}{2\hbar}d_z\left(\tilde{\rho}_{11} - \tilde{\rho}_{22}\right)\tilde{E}_z - \dfrac{1}{2\hbar}d_y\tilde{\rho}_{23}^*\tilde{E}_y - \dfrac{1}{2\hbar}d_y\tilde{\rho}_{24}^*\tilde{E}_y, \\[4pt]
\dfrac{\partial \tilde{\rho}_{13}}{\partial \tau} + \left[i\left(\omega_{13}(\tau,x) - \omega_{inc}\right) + \gamma_{13}\right]\tilde{\rho}_{13} = \dfrac{1}{2\hbar}d_y\left(\tilde{\rho}_{11} - \tilde{\rho}_{33}\right)\tilde{E}_y + \dfrac{i}{2\hbar}d_z\tilde{\rho}_{23}\tilde{E}_z - \dfrac{1}{2\hbar}d_y\tilde{\rho}_{34}^*\tilde{E}_y, \\[4pt]
\dfrac{\partial \tilde{\rho}_{14}}{\partial \tau} + \left[i\left(\omega_{14}(\tau,x) - \omega_{inc}\right) + \gamma_{14}\right]\tilde{\rho}_{14} = \dfrac{1}{2\hbar}d_y\left(\tilde{\rho}_{11} - \tilde{\rho}_{44}\right)\tilde{E}_y + \dfrac{i}{2\hbar}d_z\tilde{\rho}_{24}\tilde{E}_z - \dfrac{1}{2\hbar}d_y\tilde{\rho}_{34}\tilde{E}_y, \\[4pt]
\dfrac{\partial \tilde{\rho}_{23}}{\partial \tau} + \left[i\omega_{23}(\tau,x) + \gamma_{23}\right]\tilde{\rho}_{23} = \dfrac{i}{2\hbar}d_z\tilde{\rho}_{13}\tilde{E}_z^* + \dfrac{1}{2\hbar}d_y\tilde{\rho}_{12}^*\tilde{E}_y, \\[4pt]
\dfrac{\partial \tilde{\rho}_{24}}{\partial \tau} + \left[i\omega_{24}(\tau,x) + \gamma_{24}\right]\tilde{\rho}_{24} = \dfrac{i}{2\hbar}d_z\tilde{\rho}_{14}\tilde{E}_z^* + \dfrac{1}{2\hbar}d_y\tilde{\rho}_{12}^*\tilde{E}_y, \\[4pt]
\dfrac{\partial \tilde{\rho}_{34}}{\partial \tau} + \left[i\omega_{34}(\tau,x) + \gamma_{34}\right]\tilde{\rho}_{34} = \dfrac{1}{2\hbar}d_y\tilde{\rho}_{14}\tilde{E}_y^* + \dfrac{1}{2\hbar}d_y\tilde{\rho}_{13}^*\tilde{E}_y,
\end{cases} \tag{9}$$

In deriving Equations (9), the rotating wave approximation was also used, which implies the proximity of the carrier frequency of the high harmonic radiation to resonance with the transitions $|1\rangle \leftrightarrow |2\rangle$ and $|1\rangle \leftrightarrow |3\rangle, |4\rangle$: $|\omega_{inc} - \bar{\omega}_z| \ll \omega_{inc}$ and $|\omega_{inc} - \bar{\omega}_y| \ll \omega_{inc}$.

The system of equations (8), (9), and (2) must be supplemented with initial and boundary conditions. In the case under consideration, the influence of XUV radiation reflections from the medium boundaries can be neglected, which corresponds to the boundary condition $\vec{E}(x=0,\tau) = \vec{E}^{(inc)}(t)$. In turn, the initial conditions imply that for $\tau = 0$, of the states $|1\rangle$- $|4\rangle$, only the state $|1\rangle$ is populated: $\tilde{\rho}_{11}(x,\tau=0) = 1$ and $\tilde{\rho}_{ii}(x,\tau=0) = 0$ for $i \neq 1$. To model the amplified spontaneous emission of the active medium, it is assumed that the initial (at $\tau = 0$) values of the coherences at the inverted transitions $|1\rangle \leftrightarrow |2\rangle, |3\rangle, |4\rangle$ are random functions of the coordinate (see [33–35]), whereas the initial values of the coherences at the remaining transitions are zero.

### 3. Analytical solution and necessary conditions for the transformation of the high-harmonic polarization

We will assume that the interaction of harmonic radiation with the active medium occurs in a linear regime (the change in population differences on inverted transitions during the interaction time is insignificant). For the radiation of high harmonics considered below, detuned from resonance with the gain lines of the active medium, this approximation is fulfilled with good accuracy. In addition, we will neglect the amplified spontaneous emission and assume that the plasma is highly dispersive for the modulating IR field, namely, the condition $g_0 / \Delta K \ll 1$ is satisfied, where $g_0 = 2\pi \omega_{inc} N_{ion}^{(res)} d_z^2 / \left( \hbar c \gamma \sqrt{\varepsilon_{pl}^{(XUV)}} \right) = 4\pi \omega_{inc} N_{ion}^{(res)} d_y^2 / \left( \hbar c \gamma \sqrt{\varepsilon_{pl}^{(XUV)}} \right)$ is the amplitude gain coefficient of the resonant XUV radiation in the absence of a modulating field, $\gamma = \gamma_{12} \approx \gamma_{13} = \gamma_{14}$, and $\Delta K = \Omega \left( \sqrt{\varepsilon_{pl}^{(XUV)}} - \sqrt{\varepsilon_{pl}^{(IR)}} \right) / c$ is the addition to the wave number of the IR field, caused by the difference in the phase velocities of the high-harmonic radiation and the modulating field of the fundamental frequency. As shown in [36], when these conditions are met, the solution to the system of equations (8), (9), (2) for the resonant radiation of a seed with an arbitrary envelope has the form

$$\tilde{E}_z(x,\tau) = \int_{-\infty}^{\infty} \tilde{S}_{inc}^{(z)}(\omega) \exp\left[\tilde{G}_z(\omega) x\right] e^{-i\omega\tau} d\omega, \quad \tilde{G}_z(\omega) = \sum_{k_z=-\infty}^{\infty} \frac{g_0 J_{k_z}^2\left(P_\Omega^{(z)}\right)}{1 + i\left(\bar{\omega}_z + 2k_z\Omega - \omega - \omega_{inc}\right)/\gamma}, \quad (10a)$$

$$\tilde{E}_y(x,\tau) = \int_{-\infty}^{\infty} \tilde{S}_{inc}^{(y)}(\omega) \exp\left[\tilde{G}_y(\omega) x\right] e^{-i\omega\tau} d\omega, \quad \tilde{G}_y(\omega) = \sum_{k_y=-\infty}^{\infty} \frac{g_0 J_{k_y}^2\left(P_\Omega^{(y)}\right)}{1 + i\left(\bar{\omega}_y + 2k_y\Omega - \omega - \omega_{inc}\right)/\gamma}, \quad (10b)$$

where $P_\Omega^{(z,y)} = \Delta_{z,y}/(2\Omega)$ are the modulation indices for $z$- and $y$-polarized transitions, $J_k(x)$ is the Bessel function of the first kind of order $k$, and $\tilde{S}_{inc}^{(z,y)}(\omega)$ are the amplitude spectra of the $z$- and $y$-components of the high-frequency field at the front boundary of the medium ($x = 0$), which coincides with the field (3) due to the absence of reflections from the boundaries of the medium.

In this paper, in contrast to [27], we consider the radiation of harmonics with a finite envelope duration and, accordingly, a finite bandwidth. For all harmonics in the radiation spectrum of the seed (3), we use an envelope of the form

$$a(\tau) = \left[\theta(\tau) - \theta(\tau - \tau_{zero})\right] \sin^2\left(\pi\tau/\tau_{zero}\right), \quad (11)$$

where $\theta(\tau)$ is the Heaviside unit step function: $\theta(\tau) = 0$ for $\tau < 0$ and $\theta(\tau) = 1$ for $\tau \geq 0$; the parameter $\tau_{zero}$ determines the duration of the seed envelope from zero to zero. The amplitude spectrum corresponding to the envelope (11) has the form

$$\tilde{A}_{inc}(\omega) = \frac{1}{2\pi} \int_0^{\tau_{zero}} \sin^2(\pi\tau/\tau_{zero}) e^{i\omega\tau} d\tau = \frac{\tau_{zero}}{4\pi} \frac{\sin(\omega\tau_{zero}/2)}{(\omega\tau_{zero}/2)\left[1-(\omega\tau_{zero}/2\pi)^2\right]} e^{i\omega\tau_{zero}/2}. \quad (12)$$

The full width of such a spectrum at half-maximum of the Fourier transform is related to the duration $\tau_{zero}$ by the relation $\Delta\omega_{signal}^{(1/2)} = 4\pi/\tau_{zero}$; the spectrum width used further in the article, determined by the zeros of the Fourier transform closest to the central maximum, is twice as large: $\Delta\omega_{signal}^{(0)} = 2\Delta\omega_{signal}^{(1/2)} = 8\pi/\tau_{zero}$. The spectra of slowly varying amplitudes of the polarization components of the harmonic radiation (3) have the form

$$\tilde{S}_{inc}^{(z)}(\omega) = E_0 \cos(\varphi) \sum_{n=-N}^{N} \tilde{A}_{inc}(\omega - 2n\Omega), \quad \tilde{S}_{inc}^{(y)}(\omega) = E_0 \sin(\varphi) \sum_{n=-N}^{N} \tilde{A}_{inc}(\omega - 2n\Omega). \quad (13)$$

The combination of Eqs. (10) and (13) gives an analytical solution for the harmonic field at the output of an optically modulated neon-like active medium.

Similarly to [27], to describe the polarization state of high harmonic radiation we will use the ellipticity given by [34]

$$\sigma = \text{tg}(\chi), \quad (14а)$$

where the angle $\chi$ is determined by

$$\sin(2\chi) = \frac{2|\tilde{E}_z| \cdot |\tilde{E}_y| \sin(\vartheta)}{|\tilde{E}_z|^2 + |\tilde{E}_y|^2}, \quad (14б)$$

and $\vartheta \equiv \arg(\tilde{E}_z) - \arg(\tilde{E}_y)$ is the phase difference between the slowly varying amplitudes of the z- and y-polarization components of XUV radiation.

As follows from solution (10), the gain spectra for the z- and y-components of the high-frequency field, $\tilde{G}_{z,y}(\omega)$, are combs of induced gain lines at frequencies $\bar{\omega}_z + 2k_z\Omega$ (for the z-component of the field) and $\bar{\omega}_y + 2k_y\Omega$ (for the y-component of the field), where $k_z$ and $k_y$ are integers. The gain coefficients corresponding to them differ from the unperturbed gain coefficient by a factor of $J_{k_z}^2(P_\Omega^{(z)})$ and $J_{k_y}^2(P_\Omega^{(y)})$. Due to the fact that the characteristic coherence relaxation rate $\gamma$, which determines the half-width of the gain lines, is much lower than the frequency of the IR field, $\gamma/\Omega \ll 1$, induced gain lines with different numbers $k_{z,y}$ do not overlap in the general case.

At the same time, as shown in [26], when the following condition is met:

$$\bar{\omega}_y - \bar{\omega}_z = 2\Omega, \quad (15)$$

the gain spectrum for the y-polarized component of the high-frequency field, $\tilde{G}_y(\omega)$, is shifted as a whole relative to the gain spectrum for the z-polarized component, $\tilde{G}_z(\omega)$, by twice the frequency of the modulating field. As a result, the gain spectra for the polarization components of the high-frequency field completely overlap, and the gain line for the z-polarized field component with number $k_z$ and gain factor $J_{k_z}^2(P_\Omega^{(z)})$ coincides with the gain line for the y-polarized component with number $k_y = k_z - 1$ and gain factor $J_{k_z-1}^2(P_\Omega^{(y)})$. This allows amplifying sets of high harmonics of the IR field of arbitrary elliptical polarization, as well as controlling their ellipticity during the amplification process due to the difference in gain factors for orthogonal polarization components [26, 35]. Condition (15) can be written in an equivalent form: $P_\Omega^{(y)} - P_\Omega^{(z)} = 1$. For neon-like $Ti^{12+}$ ions, this equality is satisfied for $P_\Omega^{(y)} = 13.57$ and $P_\Omega^{(z)} = 12.57$, while the gain spectra for the z- and y-polarized field components contain about 25 induced gain lines of substantially non-zero amplitude, see Fig. 3 in [26]. As shown in [26, 38], on the left wing of the gain spectrum, for gain lines with $-14 \leq k_z = k_y + 1 \leq -6$, see Fig. 2, the gain factors for the z- and y-polarization components of the field practically coincide, which makes it possible to amplify a set of high-order harmonics of circular or elliptical polarization while maintaining their polarization state.

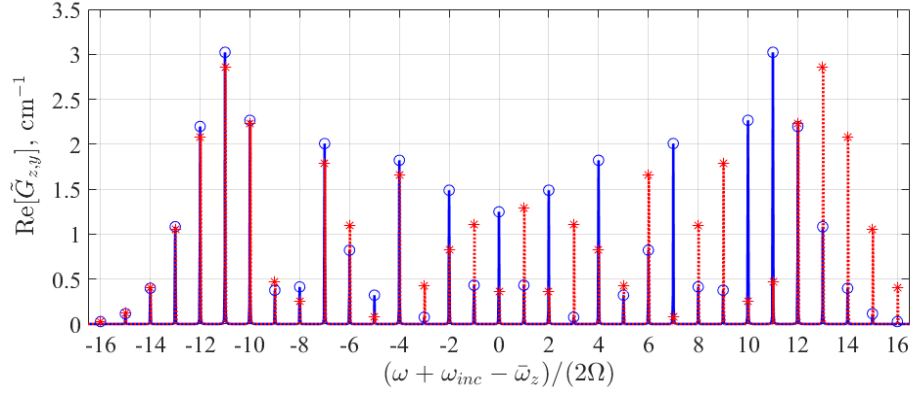

**Fig. 2.** The gain spectrum (field gain increment) of optically modulated active plasma of neon-like Ti$^{12+}$ ions (10) under conditions of Eq. (16) for $\Delta\omega_{yz}/(2\gamma) \simeq 1.9$. The blue solid line and blue circles correspond to the gain spectrum for the *z*-polarized field; the red dotted line and red stars are for the *y*-polarized field. The unperturbed gain of the medium is $g_0 = 35$ cm$^{-1}$.

In the preceding work [27], in the approximation of monochromatic harmonic fields, it was shown that the transformation of linearly polarized high harmonics into circularly polarized ones becomes possible in the case where condition (15) is not fulfilled exactly, and there is a certain detuning between the gain lines for the *z*- and *y*-polarized components of the field:

$$\bar{\omega}_y - \bar{\omega}_z = 2\Omega + \Delta\omega_{yz}. \tag{16}$$

In this case, if the value of $\Delta\omega_{yz}$ is positive and comparable with the width of the gain lines of the active medium, $2\gamma$, a spectral window appears between the $k_z$-th gain line for the *z*-polarized field and the ($k_z - 1$)-th gain line for the *y*-polarized field, in which the *z*-polarized component of the field experiences strong positive resonant dispersion, whereas the *y*-polarized component experiences strong negative dispersion. As a result, with the optimal choice of (i) the frequency of the high harmonic between the gain lines for the *z*- and *y*-polarized field components and (ii) the orientation of its polarization relative to the polarization of the modulating field (angle $\varphi$, see Fig. 1), at a certain thickness of the active medium, the *z*- and *y*-polarized components of the harmonic field acquire a phase difference of $\pi/2$, while their amplitudes are equalized, which corresponds to a change in the polarization of the field from linear to circular. In particular, for gain lines with $-14 \leq k_z \leq -6$ (see Fig. 2), neglecting the difference in peak gain factors for the *z*- and *y*-polarized field components, the harmonic frequency must be chosen exactly in the middle between the *z*- and *y*-polarized gain lines, and its polarization must be oriented at an angle $\varphi = \pi/4$ to the modulating field polarization. Note that the efficiency of such polarization conversion exceeds 100%, since the harmonic radiation is amplified during propagation in the medium.

### 4. Transformation of polarization of the field of a single high-order harmonic

Let us analyze the optimal conditions for the transformation of the polarization of the radiation of a single high-order harmonic, taking into account its finite bandwidth (finite duration of the envelope). In this case, as for the monochromatic radiation of the harmonic [27], it is advisable to use any of the gain lines with $-14 \leq k_z \leq -6$ to transform the polarization. However, due to the non-uniformity of the resonance dispersion and gain within the spectral contour of the high harmonic, the transformation of the harmonic polarization will be non-uniform across the spectrum, which will lead to a non-uniform distribution of ellipticity within the high harmonic pulse in the time domain.

To analyze the resulting distortions of the polarization state of the harmonic, we consider the frequency dependences of (i) the difference in the dispersion curves for the orthogonal polarization components of the harmonic field, $D(\omega) = \text{Im}\{\tilde{G}_z\} - \text{Im}\{\tilde{G}_y\}$, as well as (ii) the gain factors for the $z$- and $y$-polarization field components, $\text{Re}\{\tilde{G}_z\}$ and $\text{Re}\{\tilde{G}_y\}$, shown in Fig. 3 for different values of the detuning $\Delta\omega_{yz}$. These dependences allow us to estimate how non-uniform the transformation of polarization of the frequency components will be within the spectral contour of the harmonic field. Figure 3(a) is plotted for $\Delta\omega_{yz}=2\gamma$, while Figs. 3(b, c, d) are plotted for $\Delta\omega_{yz}=3.46\gamma$, $3.8\gamma$, and $7.6\gamma$, respectively. The above dependences are normalized to $g_0 J_{k_z}^2$, where $J_{k_z}^2 \equiv \left[ J_{k_z}^2\left(P_\Omega^{(z)}\right) + J_{k_z-1}^2\left(P_\Omega^{(y)}\right) \right]/2$. In Fig. 3, the peak gain factors for the $z$- and $y$-polarized field components are assumed to be equal, which is a good approximation for the induced gain lines with $-14 \le k_z \le -6$ (see Fig. 2). In this case, as for monochromatic harmonic fields [27], the optimal harmonic frequency is located exactly in the middle between the gain lines for the $z$- and $y$-polarized field components, and the optimal value of the angle $\varphi$ is $\pi/4$. Note that in the calculations presented below (both numerical and analytical), the difference between $J_{k_z}^2\left(P_\Omega^{(z)}\right)$ and $J_{k_z-1}^2\left(P_\Omega^{(y)}\right)$ is taken into account.

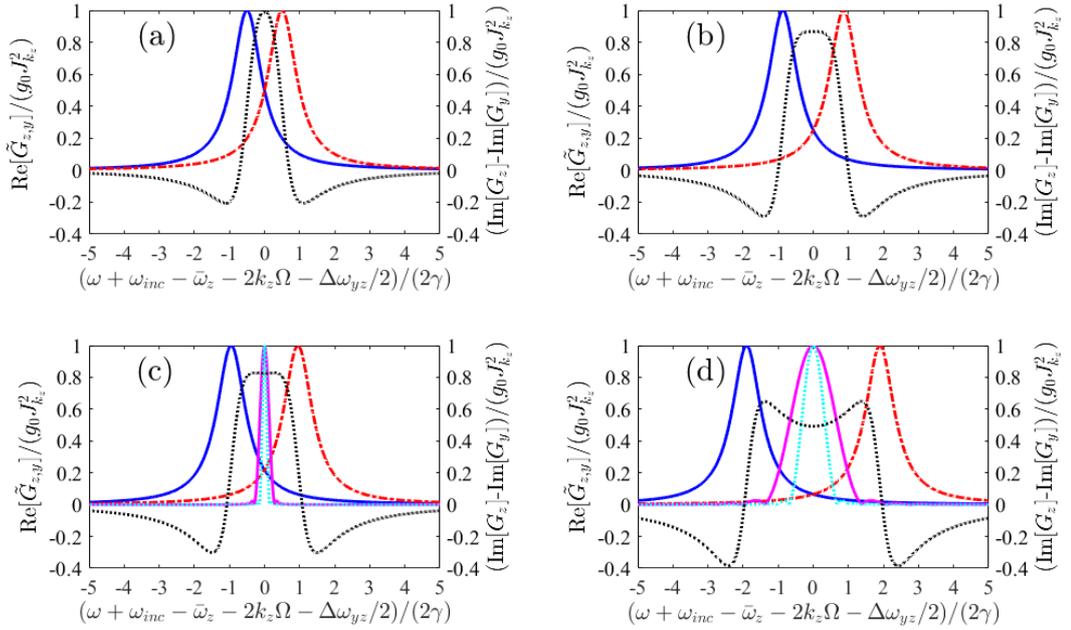

**Fig. 3.** Frequency dependences of (a) the gain (blue solid line for $z$-polarization and red dash-dotted line for $y$-polarization; left vertical axis) and (b) the difference in dispersion curves for the $z$- and $y$-polarization components of the resonant harmonic field (black dotted line, right vertical axis). Figures (a), (b), (c), and (d) correspond to $\Delta\omega_{yz}/(2\gamma) = 1$, 1.73, 1.9, and 3.8. (c) and (d) show the amplitude spectra of the resonant harmonic field (12), with the carrier frequency in the middle between the gain lines for the $z$- and $y$-polarized field. In (c), magenta corresponds to $\Delta\omega_{signal}^{(0)}/(2\gamma) = 0.568$, cyan to $\Delta\omega_{signal}^{(0)}/(2\gamma) = 0.284$, whereas in (d), purple is for $\Delta\omega_{signal}^{(0)}/(2\gamma) = 2.75$ and cyan is for $\Delta\omega_{signal}^{(0)}/(2\gamma) = 1.38$ (the width of the field spectrum at half the amplitude modulus is two times smaller). The figure is valid for any gain line with number $-14 \le k_z \le -6$.

When using the chosen normalization, the peak values of the gain factors are $\max\{\text{Re}[\tilde{G}_z]/(g_0 J_{k_z}^2)\} = \max\{\text{Re}[\tilde{G}_y]/(g_0 J_{k_z}^2)\} = 1$. The absolute maximum of the normalized difference of the dispersion curves $D(\omega)/(g_0 J_{k_z}^2)$ is also equal to 1 and is achieved at $\Delta\omega_{yz}=2\gamma$ at a frequency located exactly in the middle between the gain lines, see Fig. 3(a). Thus, the value $\Delta\omega_{yz}=2\gamma$ ensures the most efficient transformation of the harmonic field polarization in the case where the width of its spectrum is much smaller than the width of the gain line [27]. In addition, at $\Delta\omega_{yz}=2\gamma$ the field of the harmonic tuned to the maximum of the function $D(\omega)$ is effectively amplified (the normalized gain is 1/2). However, for the transformation of the field of the harmonic with a spectrum width comparable to $\gamma$, the value $\Delta\omega_{yz}=2\gamma$ is unsuitable, since in this case the difference in the dispersion curves and the gain factors for the orthogonal polarization components of the field differ significantly for different spectral components within the spectral contour of the harmonic field. In this case, it is preferable to use large values of the gain line detuning $\Delta\omega_{yz}$ (see Fig. 3(c, d)).

In connection with the above, the question arises as to what is the optimal value of the detuning $\Delta\omega_{yz}$ for a given bandwidth of the harmonic field. To answer this question, we will analyze the dependence of the normalized difference of the dispersion curves for the $z$- and $y$-polarization components of the field, $D(\omega)/(g_0 J_{k_z}^2)$, on the value of $\Delta\omega_{yz}$. First of all, we note that regardless of the value of $\Delta\omega_{yz}$, the extremum of the dependence $D(\omega)$ (i.e. the point where $dD(\omega)/d\omega = 0$) is located in the middle between the gain lines for the $z$- and $y$-polarized field components. In this case, if $0 < \Delta\omega_{yz}/(2\gamma) < 1.73$, then this extremum is a maximum, if $\Delta\omega_{yz}/(2\gamma) > 1.73$, then it is a minimum, and if $\Delta\omega_{yz}/(2\gamma) \simeq 1.73$, then it is an inflection point ($d^2 D(\omega)/d\omega^2 = 0$), see Fig. 3(b). Next, we introduce the concept of a spectral window of anisotropic dispersion for the $z$- and $y$-polarization components of the field. We define this window as the distance between the nearest extrema of the function $D(\omega)$ adjacent to the central extremum. If $0 < \Delta\omega_{yz}/(2\gamma) < 1.73$, then these extrema are minima (at $\Delta\omega_{yz}/(2\gamma) = 1$, see Fig. 3(a), they are shifted from the central maximum of the dependence $D(\omega)$ by $\pm 1.10\gamma$). If $\Delta\omega_{yz}/(2\gamma) > 1.73$, then the neighboring extrema are maxima (at $\Delta\omega_{yz}/(2\gamma) = 3.8$, see Fig. 3(d), they are shifted from the central minimum of $D(\omega)$ by $\pm 1.38\gamma$). If $\Delta\omega_{yz}/(2\gamma) \simeq 1.73$ (see Fig. 3(b)), then the width of the spectral window is zero. The dependence of the spectral window width $\Delta\omega_{disp}$ on the detuning value $\Delta\omega_{yz}$ determined in this way is shown in Fig. 4(a). Due to the fact that the definition of $\Delta\omega_{disp}$ changes at $\Delta\omega_{yz}/(2\gamma) \simeq 1.73$, the dependence $\Delta\omega_{disp}(\Delta\omega_{yz})$ at this point exhibits a discontinuity. Note also that at $\Delta\omega_{yz} \to 0$ the value of $\Delta\omega_{disp}/(2\gamma)$ remains finite and tends to 1.73.

Figure 4(b) shows the dependences on the detuning $\Delta\omega_{yz}$ for (1) the peak value of the difference in the dispersion curves for the $z$- and $y$-polarized field components (red dotted curve), (2) the average value of the difference in the dispersion curves, $\text{mean}[D(\omega)]$, calculated within the spectral window $\Delta\omega_{disp}$ (blue solid curve), and (3) the root-mean-square deviation $\delta[D(\omega)]$ of the $D(\omega)$ quantity from its average value within the spectral window $\Delta\omega_{disp}$ (black dash-dotted curve). Figure 4(c) shows the dependence on $\Delta\omega_{yz}$ for the average value of the modulus of the difference in the spectral gain contours for the $z$- and $y$-polarized components, $\text{mean}[|\Delta G(\omega)|]$, $\Delta G(\omega) = \text{Re}[\tilde{G}_z(\omega)] - \text{Re}[\tilde{G}_y(\omega)]$, calculated within the spectral window $\Delta\omega_{disp}$.

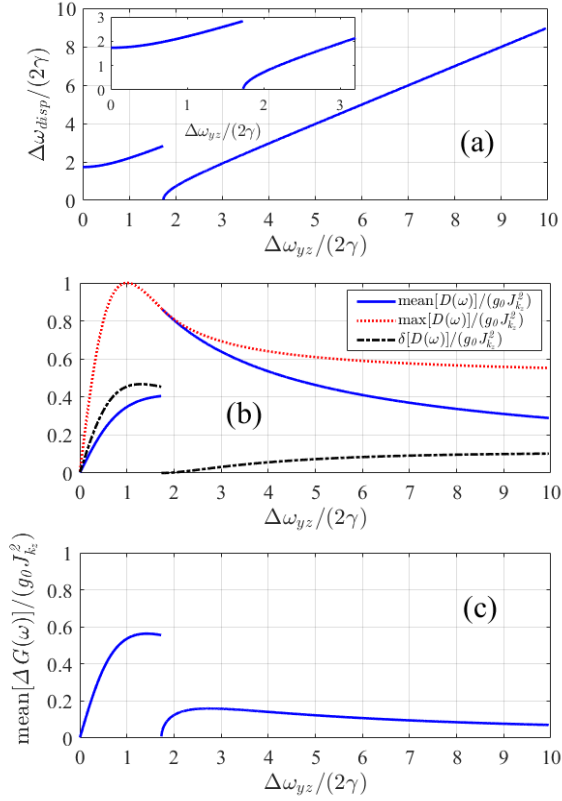

**Fig. 4.** (a) Dependence of the width of the spectral window of anisotropic dispersion $\Delta\omega_{disp}$, formed between the induced gain lines for orthogonal polarization components of the field, on their frequency detuning $\Delta\omega_{yz}$. (b) Peak and mean values, as well as the standard deviation from the mean value of the difference of dispersion curves within the spectral window $\Delta\omega_{disp}$ depending on $\Delta\omega_{yz}$. (c) Average value of the modulus of the difference in spectral gain contours for the $z$- and $y$-polarized field components within the spectral window $\Delta\omega_{disp}$ as a function of $\Delta\omega_{yz}$. The figure is valid for any gain line with number $-14 \leq k_z \leq -6$.

All the above dependencies are normalized to the value of $g_0 J_{k_z}^2$, which makes Fig. 4 valid for any gain line with number $-14 \leq k_z \leq -6$ (see Fig. 2).

Next, we reformulate the question about the optimal value of the detuning $\Delta\omega_{yz}$ as follows: for what values of $\Delta\omega_{yz}$ is the polarization transformation of the harmonic field with a spectrum width $\Delta\omega_{signal}^{(0)} = \Delta\omega_{disp}$ sufficiently effective (the required thickness of the active medium is not too large) and, at the same time, sufficiently uniform? As follows from Fig. 4(b,c), in the region $0 < \Delta\omega_{yz}/(2\gamma) < 1.73$ the standard deviation of $D(\omega)$ within the spectral window $\Delta\omega_{disp}$ exceeds its average value: $\delta[D(\omega)] > \text{mean}[D(\omega)]$, in addition, the average value of the modulus of the difference in the spectral gain contours of the $z$- and $y$-polarizations also exceeds the average value of the difference in the dispersion curves: $\text{mean}[|\Delta G(\omega)|] > \text{mean}[D(\omega)]$. This makes the interval $0 < \Delta\omega_{yz}/(2\gamma) < 1.73$ unsuitable for polarization transformation of a field with a bandwidth of $\Delta\omega_{disp}$. Indeed, in this case, for different spectral components of the harmonic field at the exit from the medium, both the phase difference and the ratio of the amplitudes of the polarization components will differ significantly, which will lead to a significant difference in their polarization.

At the same time, for $\Delta\omega_{yz}/(2\gamma) > 1.73$, the polarization transformation turns out to be much more uniform: within the spectral window width $\Delta\omega_{disp}$, the average value of $D(\omega)$ is many times greater than both its standard deviation, $\text{mean}[D(\omega)] \gg \delta[D(\omega)]$, and the average value of the modulus of the difference in the spectral gain contours, $\text{mean}[D(\omega)] \gg \text{mean}[|\Delta G(\omega)|]$. As a result, the field of a single harmonic with a bandwidth of the order of $\Delta\omega_{disp}$ has a well-defined polarization state at the output of an extended medium.

In what follows, we will consider two values of the gain line detuning for the $z$- and $y$-polarized field components: $\Delta\omega_{yz}/(2\gamma) = 1.9$ (similar to the paper [27]), see Fig. 3(c), and a value twice as large: $\Delta\omega_{yz}/(2\gamma) = 3.8$, see Fig. 3(d). At $\Delta\omega_{yz}/(2\gamma) = 1.9$, the dependence $D(\omega)$ contains a pronounced shelf, where $\delta[D(\omega)] \simeq 10^{-3} g_0 J_{k_z}^2$ and $\text{mean}[|\Delta G(\omega)|] \simeq 0.1 g_0 J_{k_z}^2$, inside the spectral window of width $\Delta\omega_{disp}/(2\gamma) = 0.568$, caused by the overlap of the dispersion curves for the $z$- and $y$-polarized field components. At $\Delta\omega_{yz}/(2\gamma) = 3.8$, the width of the spectral window increases almost fivefold: $\Delta\omega_{disp}/(2\gamma) = 2.75$. In this case, the overlap of the dispersion curves decreases, and the dependence $D(\omega)$ approaches the set of two independent resonance dispersion curves, as a result of which the value of $\delta[D(\omega)]$ increases approximately 50 times, while the value of $\text{mean}[|\Delta G(\omega)|]$ increases approximately 1.5 times in comparison with the case of $\Delta\omega_{yz}/(2\gamma) = 1.9$. Thus, an increase in the detuning $\Delta\omega_{yz}$ allows one to expand the spectrum of the converted radiation at the cost of increasing the inhomogeneity of the polarization conversion.

Next, we will analyze the spatial-temporal behavior of the ellipticity of the high harmonic radiation of the modulating field, shown in Fig. 5, which has a linear polarization with $\varphi = \pi/4$ (see Fig. 1) at the input to the medium and a carrier frequency exactly in the middle between the induced gain lines with $k_z = -11$ and $\Delta\omega_{yz}/(2\gamma) = 1.9$: $\omega_{inc} = \bar{\omega}_z + 2k_z\Omega + \Delta\omega_{yz}/2$. The spectral contour of the harmonic field is determined by Eq. (12) with $\Delta\omega_{signal}^{(0)} = \Delta\omega_{disp} = 0.568 \times (2\gamma)$. For the convenience of further discussion, we will write the solution of the wave equation (8) in the following form:

$$\vec{\tilde{E}}(x,\tau) = \vec{\tilde{E}}_{inc}(\tau) + \vec{\tilde{E}}_{scatt}(x,\tau),$$

$$\vec{\tilde{E}}_{scatt}(x,\tau) \equiv i2\pi \frac{\omega_{inc}}{c\sqrt{\varepsilon_{pl}^{(XUV)}}} \int_0^x \vec{\tilde{P}} dx'. \qquad (17)$$

Here $\vec{\tilde{E}}_{inc}(\tau)$ the slowly varying amplitude of the high-harmonic field at the front boundary of the medium, and $\vec{\tilde{E}}_{scatt}(x,\tau)$ is the slowly varying amplitude of the coherently scattered field, the source of which is the resonant polarization of the medium (4). For the modulating field considered below with a wavelength of 3.9 μm, when condition (16) is satisfied, the order of the harmonic tuned between the gain lines with $k_z = -11$ is 133, and its wavelength is 29.3 nm. Figures 5(a, b) are plotted based on the analytical solution of (10)-(13), while Figs. 5(c, d, e, f) show the results of the numerical solution of a more general system of equations (8), (9), and (2) for peak seeding radiation intensities of $10^9$ W/cm$^2$ (Figs. 5(c, d)) and $10^{10}$ W/cm$^2$ (Figs. 5(e, f)). Unlike the analytical solution, the results of the numerical calculations take into account the nonlinearity in the interaction of the harmonic radiation with the active medium, as well as the generation of amplified spontaneous emission from quantum noise. With the considered width $\Delta\omega_{signal}^{(0)}$ of the harmonic spectrum, the duration of the harmonic field pulse at the input to the medium at half the intensity maximum is $\Delta\tau_{1/2}$=1.72 ps, while the total duration of the time interval during which the incident field is different from zero is $\tau_{zero} = 4.7$ ps.

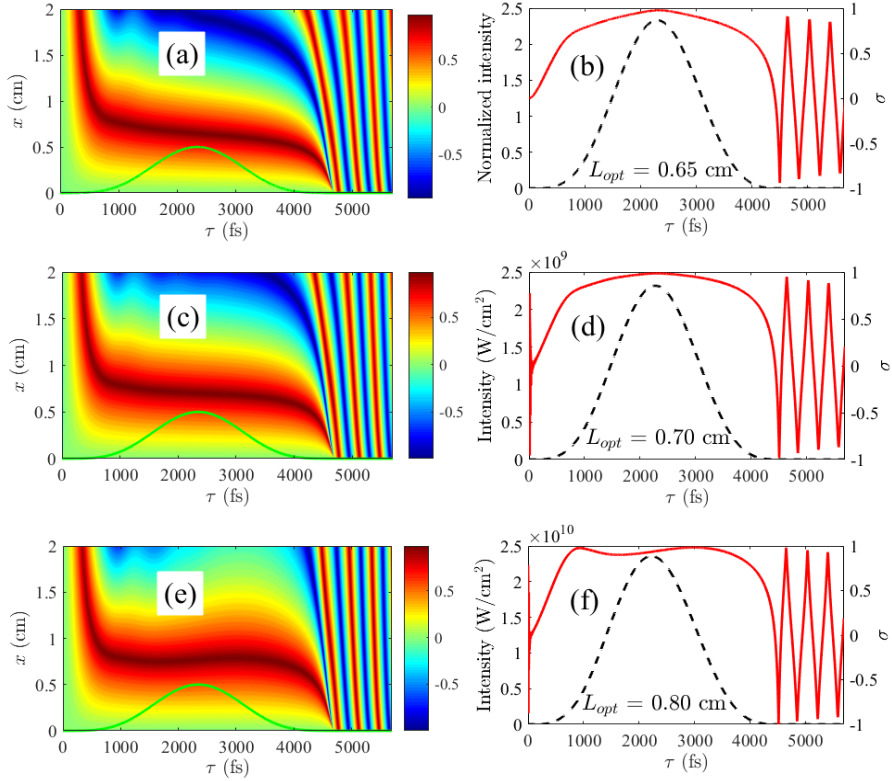

**Fig. 5.** (a,c,e) Spatiotemporal behavior of the ellipticity of the harmonic radiation with a central frequency in the middle between the gain lines for the *z*- and *y*-polarized field components with $k_z = -11$, $\Delta\omega_{yz}/(2\gamma) = 1.9$. At the input to the medium, the field is linearly polarized with $\varphi = \pi/4$; $\Delta\omega_{signal}^{(0)} = \Delta\omega_{disp} = 0.568 \times (2\gamma)$. The green solid curve in these figures shows the envelope of the harmonic field at the input to the medium. (b,d,f) Time dependences of the ellipticity (right vertical axis, red solid line) and the total intensity over polarization components (left vertical axis, black dashed line) of the harmonic field at the optimal thicknesses of the medium indicated in the corresponding figures. Figures (a) and (b) correspond to the analytical solution (10), (12), (13), whereas figures (c), (d) and (e), (f) correspond to the numerical solution of equations (8), (9), and (2) for peak intensities of seed radiation of $10^9$ W/cm² and $10^{10}$ W/cm², respectively. The time dependence of the intensity in (b) is normalized to the peak intensity of the seed field, since the behavior of the analytical solution is independent of the intensity; in (d) and (f) the intensity is given in units of W/cm². All figures are drawn for a modulating field wavelength of 3.9 μm and an intensity of $8.42 \times 10^{16}$ W/cm², for which condition (16) is satisfied at $\Delta\omega_{yz}/(2\gamma) = 1.9$. The corresponding values of the modulation indices for *z*- and *y*-polarized transitions are $P_\Omega^{(z)} \simeq 12.80$ and $P_\Omega^{(y)} \simeq 13.82$.

First of all, we will discuss the analytical solution shown in Fig. 5(a). Since the incident field is linearly polarized, at *x*=0 the field ellipticity is zero at any moment of local time τ. At initial moments of time, less than the time of establishing the resonant response of the medium, $\tau < 1/\gamma$, where $1/\gamma \approx 220$ fs [26], at the considered thicknesses of the medium, $0 \leq x \leq 2$ cm, the field ellipticity remains nearly zero, since the response of the medium does not have time to develop by this time, and the coherently scattered field is close to zero. Then, at $1/\gamma \leq \tau \leq 3/\gamma$, at the maximum considered thickness of the medium, $x = 2$ cm, the ellipticity quickly reaches a local maximum, σ = 0.832, and again decreases almost to zero. In turn, the ellipticity maximum shifts with increasing local time from $x = 2$ cm to the region $x < 1$ cm. This behavior is due to the increase in the amplitude of the resonant response of the medium (coherently scattered field), caused by (a) the establishment of the resonant response at times of the order of 1/γ and (b) the increase in the amplitude of the incident field inducing resonant polarization. As a result, at the maximum considered thickness of the medium, $x = 2$ cm, at $\tau = 328$ fs, the phase difference between the $z$- and $y$-polarized components of the harmonic field reaches $\pi/2$, which, at close amplitudes of the polarization components, corresponds to an ellipticity of the order of unity, and continues to grow, increasing to $\pi$ at $\tau = 579$ fs, which corresponds to zero ellipticity. With a further increase in local time, at $3/\gamma \leq \tau \leq \tau_{zero} - 3/\gamma$, the maximum positive ellipticity of the field is achieved at a thickness of the medium smoothly decreasing from 0.8 cm to 0.4 cm. As before, the shift of the ellipticity maximum to the region of smaller thicknesses of the medium is caused by the amplification of the coherently scattered field compared to the incident field. In the final time interval preceding the switching off of the incident field, $\tau_{zero} - 3/\gamma \leq \tau \leq \tau_{zero}$, the ellipticity maximum rapidly approaches the front boundary of the medium, $x = 0$, which is associated with a decrease in the amplitude of the incident field and, as a consequence, an increase in the relative amplitude of the coherently scattered field induced by the incident field at previous moments of time. Note that at any moment in time, with increasing thickness of the medium, the phase difference between the $z$- and $y$-polarized components of the field increases monotonically. As a result, after reaching the maximum (phase difference $\pi/2$), the ellipticity decreases to zero (phase difference $\pi$) and reaches its minimum (phase difference $3\pi/2$), corresponding to the opposite (relative to the maximum positive ellipticity) direction of rotation of the electric field vector.

Note that after the incident field pulse ends, the ellipticity in Fig. 5(a) oscillates from zero to a value close to unity; the positions of the maxima and minima of the oscillations depend weakly on the thickness of the medium. These oscillations correspond to the radiation of free induction decay (FID) of the induced polarization of the medium. Due to the presence of a detuning between the gain lines for the $z$- and $y$-polarization fields, the central frequencies of the orthogonal polarization components of the FID field differ by a value of the order of $\Delta\omega_{yz} = 3.8\gamma$. As a result, over time, a phase difference $\Delta\psi(x,\tau) \simeq \Delta\omega_{yz}\tau$ develops between them, which leads to a periodic change in the field ellipticity over time with a period $T_{osc} \simeq 2\pi/\Delta\omega_{yz} \simeq 351$ fs and does not depend on the thickness $x$ of the medium. In this case, the ellipticity at the oscillation maxima turns out to be somewhat less than unity due to the difference in the amplitudes of the $z$- and $y$-polarization components of the field, caused by a small difference in the corresponding gain coefficients.

Figures 5(c,e) show similar spatiotemporal dependences of ellipticity calculated by numerically solving equations (8), (9), and (2) for the peak seed radiation intensity of $10^9$ W/cm$^2$ and $10^{10}$ W/cm$^2$, respectively. Figures 5(c) and (a) are very similar, indicating the high accuracy of the analytical solution (10) for describing the field with moderate intensity. The main difference of Fig. 5(c) is the shift of the extrema of the spatiotemporal dependence of ellipticity to larger thicknesses of the medium. This effect increases with increasing local time $\tau$ and is caused by a decrease in the population difference at the inverted transitions of the active medium during the amplification of the harmonic field, which leads to a weakening of the resonant interaction and, as a consequence, a decrease in the phase difference between the polarization components of the harmonic field at a fixed thickness of the medium. In Fig. 5(e), the depletion of the population inversion with increasing local time is even more pronounced and leads to the fact that at the trailing edge of the field pulse, the thickness of the medium required to impart circular polarization to the field locally increases.

Note that at significantly lower intensities of incident radiation at the leading edge of the harmonic pulse, another difference between the numerical solution and the analytical one becomes noticeable: in the numerical solution, the space-time dependence of the ellipticity at the initial moments of time ($\tau \leq 3/\gamma$) turns out to be noisy due to the enhanced spontaneous radiation of the active medium.

Figures 5(b, d, f), in turn, depict the time dependences of (i) ellipticity and (ii) the total intensity of the harmonic field over the polarization components at the optimal thickness of the medium $L_{opt}$, which is determined from the condition of coincidence of the maxima of these functions in time. Figure 5(b) corresponds to the analytical solution, represented in Fig. 5(a) with $L_{opt} = 0.65$ cm, while Figs. 5(d, f) show the results of the numerical solutions, represented in Figs. 5(c), (e),

with $L_{opt} = 0.70$ and $0.80$ cm, respectively. At the optimal thickness of the medium, Figs. 5(b) and (d) practically coincide. The incident pulse is amplified without changing its shape and duration and acquires circular polarization at the maximum intensity. In this case, the radiation ellipticity becomes time-dependent; however, for the considered combination of the values of gain line detuning and width of the incident field spectrum, the change in ellipticity within the central part of the pulse is insignificant. Thus, in Fig. 5(b), at the maximum intensity, the ellipticity is 0.977, while at half-height, the ellipticity is 0.882. The main feature of Fig. 5(f) is the double-humped structure of the time dependence of ellipticity, caused by the depletion of the population inversion during the interaction of radiation with the medium (see the section of Fig. 5(e) for $x = 0.8$ cm).

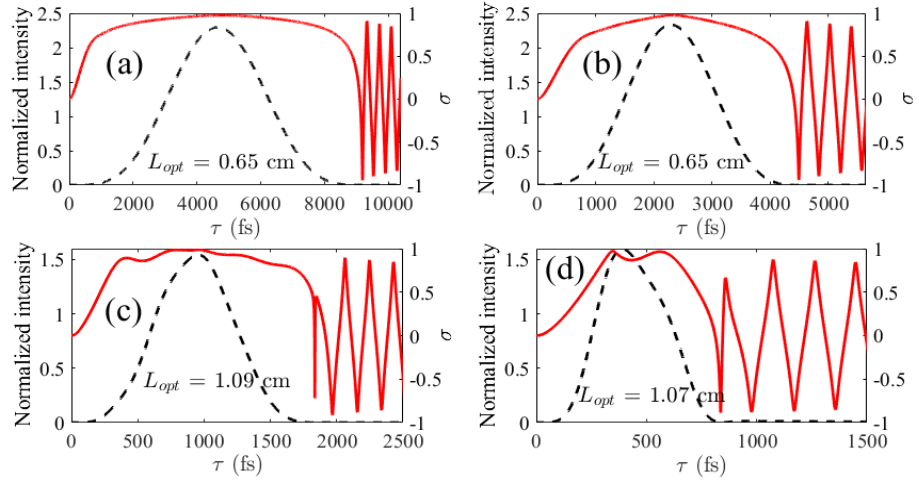

**Fig. 6.** Time dependences of the normalized intensity (left vertical axis, black dashed line) and ellipticity (right vertical axis, red solid line) of the harmonic field at the optimal thickness of the medium indicated in each figure. The intensity is normalized to the peak intensity of the harmonic field at the input to the medium. Figure (a) corresponds to $\Delta\omega_{yz}/(2\gamma) = 1.9$ and $\Delta\omega_{signal}^{(0)} = \Delta\omega_{disp}/2 = 0.284 \times (2\gamma)$; figure (b) corresponds to $\Delta\omega_{yz}/(2\gamma) = 1.9$ and $\Delta\omega_{signal}^{(0)} = \Delta\omega_{disp} = 0.568 \times (2\gamma)$; figure (c) corresponds to $\Delta\omega_{yz}/(2\gamma) = 3.8$ and $\Delta\omega_{signal}^{(0)} = \Delta\omega_{disp}/2 = 1.38 \times (2\gamma)$; figure (d) corresponds to $\Delta\omega_{yz}/(2\gamma) = 3.8$ and $\Delta\omega_{signal}^{(0)} = \Delta\omega_{disp} = 2.75 \times (2\gamma)$. All figures are drawn on the basis of the analytical solution (10). The remaining parameters of the medium and the field are the same as in Fig. 5.

Figure 6 shows the time dependences of (i) the ellipticity and (ii) the total polarization integrated intensity of the harmonic tuned between the gain lines with $k_z=-11$ on the optimal thickness of the medium, similar to those in Figs. 5(b, d, f), calculated on the basis of the analytical solution (10) for various combinations of the values of gain line detuning $\Delta\omega_{yz}$ and the incident field bandwidth $\Delta\omega_{signal}^{(0)}$. Figures 6(a) and (b) correspond to $\Delta\omega_{yz}/(2\gamma) = 1.9$, as does Fig. 5. In Fig. 6(a), the spectral width of the high harmonic contour (12) at the input to the medium is two times smaller than the width of the spectral window $\Delta\omega_{disp}$ (see Fig. 4(a)): $\Delta\omega_{signal}^{(0)} = \Delta\omega_{disp}/2 = 0.284 \times (2\gamma)$ (the corresponding spectrum is shown in Fig. 3(c) by a cyan dotted line), and Fig. 6(b) repeats Fig. 5(b) and corresponds to $\Delta\omega_{signal}^{(0)} = \Delta\omega_{disp} = 0.568 \times (2\gamma)$ (the magenta solid line in Fig. 3(c)). In turn, Figs. 6(c, d) correspond to a twofold greater detuning of the gain lines for the *z*- and *y*-polarized field components: $\Delta\omega_{yz}/(2\gamma) = 3.8$, see Fig. 3(d), which (at a fixed modulating field wavelength of 3.9 μm) is achieved by increasing the modulating field intensity from $8.42\times10^{16}$ to $8.57\times10^{16}$ W/cm². In this case, Fig. 6(c) corresponds to $\Delta\omega_{signal}^{(0)} = \Delta\omega_{disp}/2 = 1.38 \times (2\gamma)$ (the corresponding spectrum is shown in Fig. 3(d) by the cyan dotted line), and Fig. 6(d) corresponds to $\Delta\omega_{signal}^{(0)} = \Delta\omega_{disp} = 2.75 \times (2\gamma)$ (the magenta solid line in Fig. 3(d)).

Several conclusions follow from the figures presented. As the pulse of the VUV radiation of the seed is shortened, the detuning between the gain lines of the $z$- and $y$-polarizations required to transform its polarization increases. At the same time, the distortions in the time dependence of the ellipticity of the transformed radiation also increase due to the increase in the inhomogeneity in the distribution of the phase difference and the ratio of the amplitudes of the spectral components of the harmonic field within its spectral contour, see Fig. 4. However, for the active medium of neon-like Ti$^{12+}$ ions, polarization transformation remains possible up to XUV radiation durations of hundreds of femtoseconds, see Fig. 6(d). Note also that the double-humped structure in the time dependence of ellipticity in Fig. 6(d) is due to the spectral inhomogeneity of the polarization transformation, in contrast to Fig. 5(e), where such a structure arises due to the depletion of the population inversion in a strong XUV field.

## 5. Transformation of polarization of the field of a set of high-order harmonics

The proximity of the gain factors for the $z$- and $y$-polarization components of XUV radiation, characteristic of induced gain lines with $-14 \leq k_z \leq -6$ (see Fig. 2), makes it possible to transform the polarization of the set of high harmonics of the modulating field, forming a train of femto- or attosecond pulses. In this case, the difference in the peak gain factors for different $k_z$ leads to a difference in the phase incursion of the polarization components of harmonics of different orders, which, in turn, leads to non-uniformity of the polarization of the radiation of the set of high harmonics on the scale of a half-cycle of the fundamental frequency field. Nevertheless, as was shown in [27] based on a simplified mathematical model, the optimal choice of the medium thickness allows achieving circular polarization of radiation ($\sigma = \pm 1$) in the vicinity of the intensity maxima of subfemtosecond pulses. In this paper, we investigated this possibility based on a more general model that takes into account the finite duration (finite width of the spectral contour) of the harmonic field, nonlinearity in the interaction of radiation with matter, amplified spontaneous emission of the active medium, and the difference in the amplitudes of the induced gain lines for orthogonal polarization components of the field.

Figure 7 shows the result of the polarization transformation of a set of 7 harmonics (from the 127th to the 139th) of the modulating field with a wavelength of 3.9 μm, close to resonance with the induced gain lines with $-14 \leq k_z \leq -8$ (see Fig. 2), and forming at the input to the medium a train of pulses with a duration of 0.8 fs and a repetition period of 6.5 fs at a central wavelength of 29.3 nm. The central frequency of each of the harmonics is located in the middle between the corresponding gain lines for the $z$- and $y$-polarization components. The detuning between the gain lines for the $z$- and $y$-polarization components of the field is assumed to be equal to $\Delta \omega_{yz}/(2\gamma) = 1.9$, similar to Fig. 5. At the input to the medium, all harmonics are linearly polarized at an angle $\varphi = \pi/4$ to the polarization of the modulating field, i.e. to the $z$-axis (see Fig. 1), and their amplitudes and phases are equal. The peak intensity of the field of each harmonic is $(1/7) \times 10^9$ W/cm$^2$, while the peak intensity of the pulse train at the input to the medium is $7 \times 10^9$ W/cm$^2$. The envelope of the pulse train with respect to the field is determined by Eq. (11) with $\tau_{zero} = 4.7$ ps, which corresponds to the spectral contour of each harmonic of the form (12) with a width equal to the width of the spectral window of anisotropic dispersion: $\Delta\omega_{signal}^{(0)} = \Delta\omega_{disp} = 0.568 \times (2\gamma)$, see Fig. 4(a). Figures 7 (a, b, c) are drawn based on the numerical solution of equations (2), (8), (9), while Fig. 7(d) is based on the analytical solution (10) in the approximation of monochromatic fields, similar to [27].

Figure 7(a) shows the spatiotemporal behavior of the radiation ellipticity of the considered set of seven harmonics at the intensity maxima of subfemtosecond pulses. In general, Fig. 7(a) is similar to Fig. 5(c) with the only difference that the maximum ellipticity is achieved at a greater thickness of the medium, which is due to the decrease in the amplitudes of the induced gain lines with increasing separation from the gain line with $k_z = -11$, see Fig. 2. As can be seen from Fig. 7(b), for the considered parameter values, the transformation of the radiation polarization is quite uniform in time on the scale of the duration of its envelope, namely, at the optimal thickness of the medium, the ellipticity of the radiation at the maxima of subfemtosecond pulses is close to unity at all moments of time for which the peak intensity of the pulses differs significantly from zero. In turn, Fig. 7 (c) and (d) allow us to draw the following conclusions. Firstly, on the scale of the pulse repetition period (half-cycle of the modulating field) the polarization of the radiation is noticeably non-uniform, namely, at the optimal thickness of the medium the ellipticity changes from unity at the pulse maxima to negative values at their pedestal. Secondly, the analytical solution derived in the monochromatic field approximation, Fig. 7(d), describes quite well the results of the numerical calculation shown in Fig. 7(c). The difference in the values of the optimal thickness of the medium is due to the depletion of the population inversion at the inverted transition of the active medium, which is not taken into account in the analytical solution. Thirdly, despite the difference in the gain factors for harmonics of different orders, the duration of the pulses they generate has changed insignificantly, from 0.8 fs to 0.92 fs.

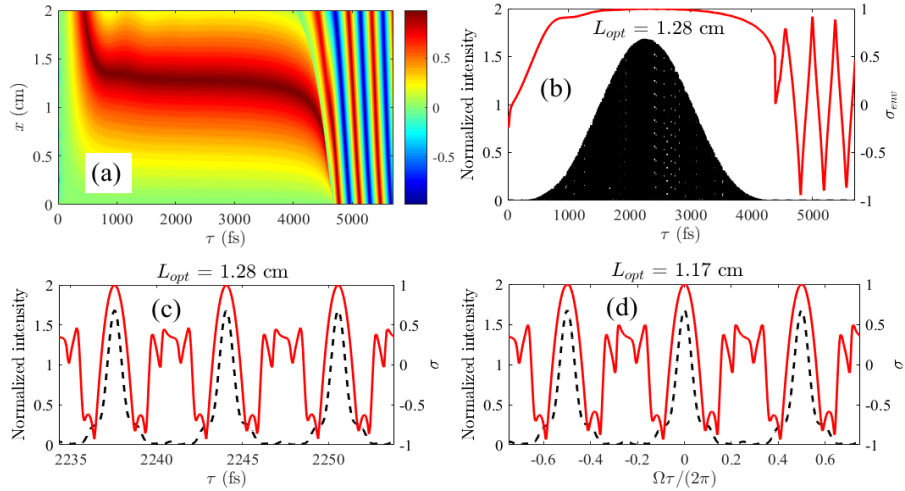

**Fig. 7.** (a) Spatiotemporal behavior of ellipticity in the maxima of subfemtosecond pulses formed by a set of 7 in-phase harmonics. At the input to the medium, the harmonics are linearly polarized ($\varphi = \pi/4$), their amplitudes are equal, and the frequencies are tuned in the middle between the gain lines for the $z$- and $y$-polarized field components with $-14 \leq k_z \leq -8$ and $\Delta\omega_{yz}/(2\gamma) = 1.9$. (b) Time dependences of the peak ellipticity (right vertical axis, red solid line) and normalized pulse intensity (left vertical axis, black dashed line) at the optimal medium thickness indicated in the figure. Intensity is normalized to the peak intensity of the incident field. (c) Same as in (b), but in a time window of 1.5 cycles of the modulating field in the vicinity of the maximum of the pulse train envelope. Figures (a)-(c) are drawn based on the numerical solution of system (2), (8), (9) for $\Delta\omega_{signal}^{(0)} = \Delta\omega_{disp} = 0.568 \times (2\gamma)$ and the peak intensity of the pulse train at the input to the medium $7 \times 10^9$ W/cm$^2$. (d) Same as in Fig. (c), but based on the analytical solution (10) without taking into account the finite width of the harmonic spectral lines (similar to [27]) at the corresponding optimal thickness of the medium. The figure corresponds to a modulating field with a wavelength of 3.9 μm and an intensity of $8.42 \times 10^{16}$ W/cm$^2$. The values of the modulation indices for $z$- and $y$-polarized transitions are $P_\Omega^{(z)} \simeq 12.80$ and $P_\Omega^{(y)} \simeq 13.82$.

Finally, we will analyze the spatial dependence of the ellipticity of the radiation of a set of three, five, and seven harmonics at the maximum of the envelope of the pulse train they form, calculated on the basis of the analytical solution (10) using the monochromatic field approximation and without it. This comparison is shown in Fig. 8. The solid lines are drawn taking into account the finite bandwidth of each of the harmonics; the dotted lines correspond to the monochromatic field approximation [27]. Figure 8(a) is drawn for the detuning of the gain lines $\Delta\omega_{yz}/(2\gamma) = 1.9$ and the width of the spectrum (12) of the field of each of the harmonics $\Delta\omega_{signal}^{(0)}/(2\gamma) = 0.568$ equal to the width of the corresponding spectral window $\Delta\omega_{disp}$, see Fig. 4(a); Figure 8(b) corresponds to $\Delta\omega_{yz}/(2\gamma) = 3.8$ and $\Delta\omega_{signal}^{(0)} = \Delta\omega_{disp} = 2.75\times(2\gamma)$. The following conclusions can be drawn from Fig. 8. Firstly, with an increase in the number of harmonics, the thickness of the medium required to impart circular polarization to the field increases. Secondly, taking into account the finite width of the harmonic spectrum is not crucial in the case shown in Fig. 8(a), but is important for the case in Fig. 8(b), which is explained by the significant inhomogeneity of the resonant dispersion and gain of the medium within the width of the field spectrum in the latter case, see Figs. 4(b,c). Note also that, taking into account the finite width of the harmonic field spectrum, the optimal thickness of the medium turns out to be smaller than in the monochromatic field approximation, see Fig. 8(b). This is explained by the fact that for the parameters of Fig. 8(b) at the central frequencies of the harmonics, the difference in the dispersion curves for the orthogonal polarization components of the field reaches a minimum, see Fig. 3(d), and the phase difference between the $z$- and $y$-polarized components of the field at the central frequencies of the harmonics turns out to be smaller than for the detuned frequency components.

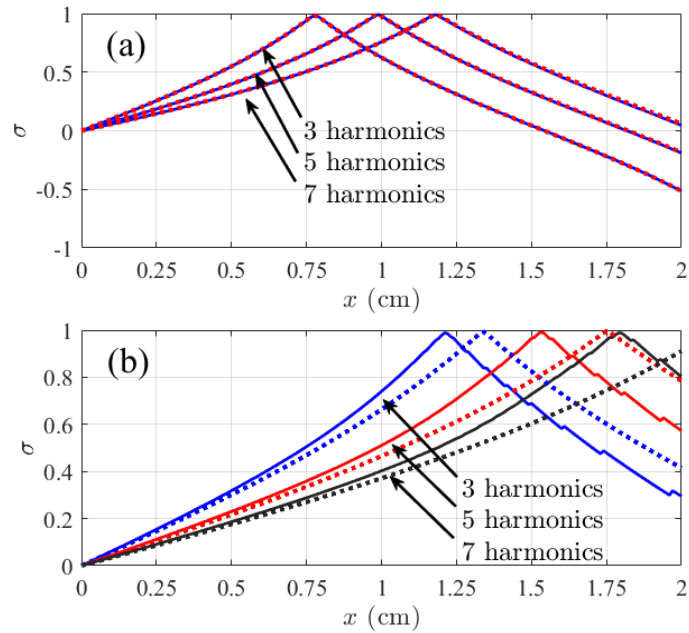

**Fig. 8.** Spatial dependence of the ellipticity of the field of a set of three, five, and seven high harmonics at the maximum of the envelope of the pulse train formed by them, calculated on the basis of the analytical solution (10). Solid lines are drawn taking into account the finite bandwidth of each of the harmonics; dotted lines correspond to the monochromatic field approximation. Figure (a) is drawn for $\Delta\omega_{yz}/(2\gamma) = 1.9$ and $\Delta\omega_{signal}^{(0)} = \Delta\omega_{disp} = 0.568\times(2\gamma)$, the modulating field has a wavelength of 3.9 μm and an intensity of $8.42\times10^{16}$ W/cm$^2$, which corresponds to $P_\Omega^{(z)} \simeq 12.80$ and $P_\Omega^{(y)} \simeq 13.82$. Figure (b) corresponds to $\Delta\omega_{yz}/(2\gamma) = 3.8$ and $\Delta\omega_{signal}^{(0)} = \Delta\omega_{disp} = 2.75\times(2\gamma)$, a modulating field with a wavelength of 3.9 μm, and an intensity of $8.57\times10^{16}$ W/cm$^2$, while $P_\Omega^{(z)} \simeq 13.03$ and $P_\Omega^{(y)} \simeq 14.07$.

### 6. Conclusion

In this paper, we investigated the possibility of transforming the polarization of high harmonics of the IR field from linear to circular during propagation in the active medium of a plasma-based X-ray laser modulated by a linearly polarized IR field of fundamental frequency. This transformation is possible due to (i) the formation of quasi-energy levels of the "resonant ions + field" system and a comb of corresponding resonances, separated from the frequency of the inverted transition of the active medium by an even number of modulation frequencies and (ii) the anisotropy of the resonant dispersion of the active medium for the polarization components of the harmonic radiation, parallel and orthogonal to the polarization of the modulating field.

The results of the conducted study can be summarized as follows. The method proposed in [27] for converting linearly polarized radiation of high harmonics into circularly polarized radiation in an optically modulated active medium of a plasma-based X-ray laser is robust with respect (i) to variation in the duration and intensity of harmonic radiation, (ii) to the presence of amplified spontaneous radiation of the active medium, and (iii) to differences in the peak gains for orthogonal polarization components of the field. The minimum duration of the harmonic radiation envelope for which the considered polarization conversion method is applicable is of the order of establishment time of the resonant response of the medium (the transverse relaxation time at the inverted transition). For the active medium of neon-like $Ti^{12+}$ ions, this is about 200 fs. At the medium output, the ellipticity of the harmonic radiation depends on time and, under optimal conditions, reaches its maximum in the vicinity of the maximum of the field intensity envelope. With the shortening of the envelope (spectral broadening) of each harmonic, the temporal non-uniformity of the polarization of the converted radiation increases, caused by the spectral non-uniformity of the resonant dispersion and gain of the active medium. At the same time, the value of the detuning between the induced gain lines for the orthogonal polarization components of the harmonic field, necessary for the transformation of its polarization, also increases.

In a low-intensity harmonic field, the amplified spontaneous emission of the active medium results in noise in the time dependence of the ellipticity at the leading edge of the converted pulse. At the same time, in a strong harmonic field, the depletion of the population inversion of the active medium results in an increase in the thickness of the medium required to achieve circular polarization.

The method under consideration allows transforming the polarization of both a single harmonic and a set of high-order harmonics. In the latter case, the temporal non-uniformity of the harmonic field polarization on the scale of its envelope duration is supplemented by non-uniformity on the scale of the half-cycle of the modulating field, caused by the difference in the amplitudes of the induced gain lines of different orders. However, in the case of neon-like $Ti^{12+}$ ions modulated by an IR field with a wavelength of 3.9 μm, it is possible to transform the polarization of a set of 3, 5, and 7 harmonics of the modulating IR field in the sense that at the optimal thickness of the medium in the vicinity of the maximum of the pulse train, the ellipticity of the total field of harmonics approaches unity at the maxima of the subfemtosecond pulses they form.

**Acknowledgments**

The work was supported by the Ministry of Science and Higher Education of the Russian Federation (Agreement No. 075-15-2020-906, Center of Excellence "Center of Photonics"). V.A. Antonov acknowledges the support from the Foundation for the Development of Theoretical Physics and Mathematics "Basis", grant No. 24-1-2-43-1.

[1] G. Schütz, W. Wagner, W. Wilhelm, P. Kienle, R. Zeller, R. Frahm, and G. Materlik, Absorption of circularly polarized x rays in iron, *Phys. Rev. Lett.* **58**, 737 (1987).

[2] C.T. Chen, F. Sette, Y. Ma, and S. Modesti, Soft-x-ray magnetic circular dichroism at the $L_{2,3}$ edges of nickel, *Phys. Rev. B* **42**, 7262 (1990).

[3] V. Schmidt, Photoionization of atoms using synchrotron radiation, *Rep. Prog. Phys.* **55,** 1483 (1992).

[4] N. Böwering, T. Lischke, B. Schmidtke, N. Müller, T. Khalil, and U. Heinzmann, Asymmetry in photoelectron emission from chiral molecules induced by circularly polarized light, *Phys. Rev. Lett.* **86**, 1187 (2001).

[5] A. Ferré, C. Handschin, M. Dumergue, F. Burgy, A. Comby, D. Descamps, B. Fabre, G.A. Garcia, R. Géneaux, L. Merceron, E. Mével, L. Nahon, S. Petit, B. Pons, D. Staedter, S. Weber, T. Ruchon, V. Blanchet, and Y. Mairesse, A table-top ultrashort light source in the extreme ultraviolet for circular dichroism experiments, *Nat. Photonics* **9**, 93 (2015).

[6] G. Lambert, B. Vodungbo, J. Gautier, B. Mahieu, V. Malka, S. Sebban, P. Zeitoun, J. Luning, J. Perron, A. Andreev, S. Stremoukhov, F. Ardana-Lamas, A. Dax, C.P. Hauri, A. Sardinha, and M. Fajardo, Towards enabling femtosecond helicity-dependent spectroscopy with high-harmonic sources, *Nat. Commun.* **6**, 6167 (2015).

[7] O. Kfir, P. Grychtol, E. Turgut, R. Knut, D. Zusin, D. Popmintchev, T. Popmintchev, H. Nembach, J.M. Shaw, A. Fleischer, H. Kapteyn, M. Murnane, and O. Cohen, Generation of bright phase-matched circularly-polarized extreme ultraviolet high harmonics, *Nat. Photonics* **9**, 99 (2015).

[8] T. Fan, P. Grychtol, R. Knut, C. Hernández-Garsía, D.D. Hickstein, D. Zusin, C. Gentry, F.J. Dollar, C.A. Mancuso, C.W. Hogle, O. Kfir, D. Legut, K. Carva, J.L. Ellis, K.M. Dorney, C. Chen, O.G. Shpyrko, E.E. Fullerton, O. Cohen, P.M. Oppeneer, D.B. Milošević, A. Becker, A.A. Jaroń-Becker, T. Popmintchev, M.M. Murnane, and H.C. Kapteyn, Bright circularly polarized soft X-ray high harmonics for X-ray magnetic circular dichroism, *Proc. Nat. Acad. Sci. USA* **112**, 14206 (2015).

[9] S. Sasaki, Analyses for a planar variably-polarizing undulator, *Nucl. Instrum. Methods Phys. Res. A* **347**, 83 (1994).

[10] E. Allaria, D. Castronovo, P. Cinquegrana et al., Two-stage seeded soft-X-ray free-electron laser, *Nat. Photonics* **7**, 913 (2013).

[11] E.A. Schneidmiller and M.V. Yurkov, Obtaining high degree of circular polarization at x-ray free electron lasers via a reverse undulator taper, *Phys. Rev. ST Accel. Beams* **16,** 110702 (2013).

[12] E. Ferrari, E. Allaria, J. Buck, G. De Ninno, B. Diviacco, D. Gauthier, L. Giannessi, L. Glaser, Z. Huang, M. Ilchen, G. Lambert, A.A. Lutman, B. Mahieu, G. Penco, C. Spezzani, and J. Viefhaus, Single shot polarization characterization of XUV FEL pulses from crossed polarized undulators, *Sci. Rep.* **5**, 13531 (2015).

[13] A.A. Lutman, J.P. MacArthur, M. Ilchen et al., Polarization control in an X-ray free-electron laser, *Nat. Photonics* **10**, 468 (2016).

[14] F. Krausz and M. Ivanov, Attosecond physics, Rev. Mod. Phys. **81**, 163 (2009).

[15] L. Young, K. Ueda, M. Gühr et al., Roadmap of ultrafast x-ray atomic and molecular physics, *J. Phys. B* **51**, 032003 (2018).

[16] R. Schoenlein, T. Elsaesser, K. Holldack, Z. Huang, H. Kapteyn, M. Murnane, and M. Woerner, Recent advances in ultrafast X-ray sources, *Philos. Trans. R. Soc. A* **377**, 20180384 (2019).

[17] K.S. Budil, P. Salières, A. L'Huillier, T. Ditmire, and M.D. Perry, Influence of ellipticity on harmonic generation, *Phys. Rev. A* **48**, R3437 (1993).

[18] P. Dietrich, N.H. Burnett, M. Ivanov, and P.B. Corkum, High-harmonic generation and correlated two-electron multiphoton ionization with elliptically polarized light, *Phys. Rev. A* **50**, R3585 (1994).

[19] C. Zhai, R. Shao, P. Lan, B. Wang, Y. Zhang, H. Yuan, S.M. Njoroge, L. He, and P. Lu, Ellipticity control of high-order harmonic generation with nearly orthogonal two-color laser fields, *Phys. Rev. A* **101**, 053407 (2020).

[20] A. Fleischer, O. Kfir, T. Diskin, P. Sidorenko, and O. Cohen, Spin angular momentum and tunable polarization in high-harmonic generation, *Nat. Photonics* **8**, 543 (2014).

[21] E. Skantzakis, S. Chatziathanasiou, P.A. Carpeggiani, G. Sansone, A. Nayak, D. Gray, P. Tzallas, D. Charalambidis, E. Hertz, and O. Faucher, Polarization shaping of high-order harmonics in laser-aligned molecules, *Sci. Rep.* **6**, 39295 (2016).

[22] X. Zhou, R. Lock, N. Wagner, W. Li, H.C. Kapteyn, and M.M. Murnane, Elliptically polarized high-order harmonic emission from molecules in linearly polarized laser fields, *Phys. Rev. Lett.* **102**, 073902 (2009).

[23] B. Vodungbo, A.B. Sardinha, J. Gautier, G. Lambert, C. Valentin, M. Lozano, G. Iaquaniello, F. Delmotte, S. Sebban, J. Lüning, and P. Zeitoun, Polarization control of high order harmonics in the EUV photon energy range, *Opt. Express* **19**, 4346 (2011).

[24] J. Schmidt, A. Guggenmos, M. Hofstetter, S.H. Chew, and U. Kleineberg, Generation of circularly polarized high harmonic radiation using a transmission multilayer quarter waveplate, *Opt. Express* **23**, 33564 (2015).

[25] A. Depresseux, E. Oliva, J. Gautier, F. Tissandier, G. Lambert, B. Vodungbo, J-P. Goddet, A. Tafzi, J. Nejdl, M. Kozlova, G. Maynard, H. T. Kim, K. Ta Phuoc, A. Rousse, P. Zeitoun, and S. Sebban, Demonstration of a circularly polarized plasma-based soft-x-ray laser, *Phys. Rev. Lett.* **115**, 083901 (2015).

[26] I.R. Khairulin, V.A. Antonov, M.Yu. Ryabikin, M.A. Berrill, V.N. Shlyaptsev, J.J. Rocca, and O. Kocharovskaya, Amplification of elliptically polarized sub-femtosecond pulses in neon-like X-ray laser modulated by an IR field, *Sci. Rep.* **12**, 6204 (2022).

[27] I.R. Khairulin, V.A. Antonov, and M.Yu. Ryabikin, On the possibility of converting linearly polarized attosecond pulses of high harmonics into circularly polarized pulses with an increase in the energy in an optically modulated neon-like active medium of an X-ray plasma laser, *JETP Letters* **117**, 652 (2023).

[28] P.V. Nickles, V.N. Shlyaptsev, M. Kalachnikov, M. Schnürer, I. Will, and W. Sandner, Short pulse x-ray laser at 32.6 nm based on transient gain in ne-like titanium, *Phys. Rev. Lett.* **78**, 2748 (1997).

[29] D. Alessi, B.M. Luther, Y. Wang, M.A. Larotonda, M. Berrill, and J.J. Rocca, High repetition rate operation of saturated table-top soft x-ray lasers in transitions of neon-like ions near 30 nm, *Opt. Express* **13**, 2093 (2005).

[30] M. Chini, B. Zhao, H. Wang, Y. Cheng, S.X. Hu, and Z. Chang, Subcycle ac Stark shift of helium excited states probed with isolated attosecond pulses, *Phys. Rev. Lett.* **109**, 073601 (2012).

[31] M.F. Gu, The flexible atomic code, *Can. J. Phys.* **86**, 675 (2008).

[32] V.S. Popov, Tunnel and multiphoton ionization of atoms and ions in a strong laser field (Keldysh theory), *Physics – Uspekhi* **47**, 855 (2004).

[33] R. Glauber and F. Haake, The initiation of superfluorescence, *Phys. Lett. A* **68**, 29 (1978).

[34] F. Haake, H. King, G. Schröder, J. Haus, R. Glauber, and F. Hopf, Macroscopic quantum fluctuations in superfluorescence, *Phys. Rev. Lett.* **42**, 1740 (1979).

[35] M. Gross and S. Haroche, Superradiance: An essay on the theory of collective spontaneous emission, *Phys. Rep.* **93**, 301 (1982).

[36] I.R. Khairulin, V.A. Antonov, M.Yu. Ryabikin, and O. Kocharovskaya, Enhanced amplification of attosecond pulses in a hydrogen-like plasma-based x-ray laser modulated by an infrared field at the second harmonic of fundamental frequency, *Photonics* **9**, 51 (2022).

[37] M. Born and E. Wolf, Principles of Optics (Pergamon Press, New York, 1964).

[38] V.A. Antonov, I.R. Khairulin, M.Yu. Ryabikin, M.A. Berrill, V.N. Shlyaptsev, J.J. Rocca, and O. Kocharovskaya, Amplification and ellipticity enhancement of high-order harmonics in a neonlike x-ray laser dressed by an IR field, *Phys. Rev. A* **107**, 063511 (2023).